\documentclass[]{interact}
\usepackage{epstopdf}% To incorporate .eps illustrations using PDFLaTeX, etc.
\usepackage[caption=false]{subfig}% Support for small, `sub' figures and tables

\usepackage[numbers,sort&compress]{natbib}% Citation support using natbib.sty

\renewcommand\bibfont{\fontsize{10}{12}\selectfont}% Bibliography support using natbib.sty
\makeatletter% @ becomes a letter
\def\NAT@def@citea{\def\@citea{\NAT@separator}}% Suppress spaces between citations using natbib.sty
\makeatother% @ becomes a symbol again

\theoremstyle{plain}% Theorem-like structures provided by amsthm.sty

\theoremstyle{definition}

\theoremstyle{remark}

\begin{document}

\articletype{ARTICLE TEMPLATE}% Specify the article type or omit as appropriate

\title{Classical Mode Dynamics for Trapped Ion Diagnostics}

\author{
\name{Itzal D.U. Terrazas and Daniel F.V. James}
\affil{Department of Physics, University of Toronto, 
60 St. George St., Toronto, Ontario, Canada M5S 1A7}
}

\maketitle

\begin{abstract}
In this paper we consider two problems in diagnostics of trapped ion crystals in which an analysis of the ions' collective oscillatory motion yield potentially useful results. When one of the ions in a linear crystal undergoes a collision, observation of the subsequent motion allows one to deduce the identity of which ion sustained the collision.  When a linear ion crystal is formed with a dark impurity ion, analysis of the ions' motion can identify the mass (and thus give an important clue to the species) of the impurity.
\end{abstract}


\section{Introduction}
When trapped atomic ions are confined in an effective three-dimensional anisotropic harmonic well (formed by radio-frequency quadrupole fields) and laser-cooled will form crystals\cite{ghosh_2007}. In these  crystals, each ion is strongly localized about one particular equilibrium position.  In conditions of strong anisotropy, these positions fall in a line along the axis of weakest trapping potential\cite{james_1998}.  Even at very low temperature, the ions undergo oscillations about their equilibrium positions, these oscillations are strongly coupled, due to the mutual Coulomb repulsion.  At very low temperatures, the fluctuations of the ions' positions must be considered as a quantum effect.

Unfortunately, ion traps have limitations, one of them being that they are set up in imperfect vacuum chambers. Ideally, the ions would be trapped in a perfect chamber without any foreign ions present, however, this is not possible. The ions can interact with gases in the chamber when lasers are being used to trap, there can be chemical reactions between the ion and the gas. One or more ions then are different from the others. This affects features of the trap, like the amplitude of the ion motion and the frequency of oscillation.

There are two interactions that can occur with the other ions in the trap, the foreign ions can collide against the trapped ions or they can be accidentally trapped. In the case of a collision there is a change in the frequency of the trap, called a collisional frequency shift (CFS) \cite{vutha_2017}. CFS was investigated for trapped-ion clocks and a fixed expression for calculating it was obtained. This value can be used to determine what the collision site is in order to remove the ion that was affected only instead of trying to restart another trap. In the latter case, the sympathetic cooling of ion motion has been a subject of research \cite{rugango_2015,kimura_2011}. Chemical reactions containing ions can be studied using ion traps, however, the mass of the product ions needs to be measured in order to get the final results. This is done using mass spectrometry\cite{fan_2021}, which is a difficult and lengthy process. In this paper results are obtained, using known and measurable quantities, to determine the location of a collision and the mass of a different ion in the trap. 

\section{Trapped ion Dynamics}
In this section we will briefly reprise the theoretical description of trapped ion dynamics; more detailed description can be found on ref.\cite{james_1998}. Let the position of the $m$-th ion in the linear chain be $x_m(t)$, where the ions are numbered from left to right, so that if $n>m$ we can assume $x_n>x_m$.  The mean position of each ion is $x_m^0$, so the position can be written as
\begin{equation}\label{position}
    x_m(t)=x_m^0+q_m(t),
\end{equation}
where $q_m(t)$ is the displacement of the $m$-th ion from equilibrium. If the ion chain was cooled to absolute zero, and quantum fluctuations somehow negligible, the $m$-th ion would be at position $x_m^0$.  However, this is not the case, so we assume that $|q_m(t)|\ll |x_{m\pm1}^0-x_m^0|$, i.e. the ions are sufficiently cool that the possibility of any two ions swapping position can be discounted.  The kinetic energy of the ions in the trap is given by 
\begin{equation}\label{k_energy}
T = \frac{1}{2}\sum_{n=1}^{\rm N}{\rm M}\dot{q}_n^2(t),
\end{equation}
where ${\rm M}$ is the mass of each ion (we assume, initially, they all have the same mass).
The potential energy includes both the trapping potential and the coulomb interaction between the ions as follows, 
\begin{eqnarray}\label{p_energy}
    V(q_1, q_2, \ldots q_{\rm N}) &=& 
    \frac{1}{2}\kappa\sum_{n=1}^{\rm N} \left(x_m^0+q_m\right)^2\nonumber\\
    &+&\frac{e^2}{8\pi\epsilon_0}\sum_{\stackrel{\scriptstyle n,m=1}{m \neq n}}^{\rm N}
    \frac{1}{|x_m^0-x_n^0+q_m-q_n|}\,\, ,\nonumber\\
    &&
 \end{eqnarray}
where $\kappa$ is a constant determined by the shape and charge of the trapping electrodes, $e$ is the charge of the electron (we assume that each ion is singly ionized) and $\epsilon_0$ is the permittivity of the vacuum.  The natural oscillation frequency of a single ion in such a trap is $\omega_0=\sqrt{\kappa/{\rm M}}$, and the potential energy will often be seen in terms of this quantity.  However, when dealing with impurity ions, it is necessary to recall that $\omega_0$ is mass-dependent, while $\kappa$ depends solely on the charge and trapping potentials.

Substituting from eq.(\ref{position}) and using the assumption $|q_m(t)|\ll |x_{m\pm1}^0-x_m^0|$, we can simplify the potential by using a Taylor expansion:
\begin{eqnarray}
    V(q_1,...,q_n)&\approx& V_0+\sum_{n=1}^{\rm N} q_n\left[\frac{\partial V}{\partial q_n}\right]_0\nonumber\\
    &+&
    \sum_{n,m=1}^{\rm N}
    \frac{q_n q_m}{2}\left[\frac{\partial^2V}{\partial q_n\partial q_m}\right]_0+ O[q_n^3],
\end{eqnarray}
where the subscript $0$ denotes the term is evaluated at $q_1 = q_2 = \ldots =q_{\rm N}=0$. Thus $V_0= V(0, 0, \ldots 0)$, which is a constant (and which we may safely disregard, since it will have no effect on the equations of motion).  Further, since the equilibrium positions $\{x_1^0, x_2^0, \ldots x_{\rm N}^0\}$ are the positions of the ions for which the potential is a minimum, by definition $[\partial V/\partial q_m]_0=0$. 
%\begin{equation}
 %   \Bigg[\frac{\partial V}{\partial q_m}\Bigg]_0=\kappa x_n^0-\frac{e^2}{4\pi\epsilon_0}\sum_{\substack{m=1\\ n\neq m}}^{\rm N}\frac{\text{sgn}(x_m^0-x_n^0)}{(x_m^0-x_n^0)^2}
%\end{equation}
The second derivative of the potential $V(q_1, q_2, \ldots q_{\rm N})$ can be obtained by taking the derivative of eq.(\ref{p_energy}); after some algebra, we obtain:
\begin{equation}
\frac{1}{\kappa}\left[\frac{\partial^2 V}{\partial q_n \partial q_m}\right]_0
\equiv A_{nm}
=
\begin{cases}
    1+\displaystyle{\sum_{\stackrel{ k=1}{k\neq m}}^{\rm N}}
    \frac{2 \ell^3}{|x_m^0-x_k^0|^3}&m=n,\\
    &\\
   \displaystyle{ \frac{-2\ell^3}{|x_m^0-x_k^0|^3}}&m\neq n,
\end{cases}
\end{equation}
where $\ell$ is a characteristic length scale given by $\ell^3=e^2/(4\pi\epsilon_0\kappa)$, and we have introduced the dimensionless ${\rm N}\times {\rm N}$ ion coupling matrix $A_{nm}$. Thus we obtain the Lagrangian for ${\rm N}$ ions in a trap:
\begin{equation}\label{eq:Lagrangianq_m}
    L = \frac{1}{2}{\rm M}\sum_{n=1}^{\rm N} \dot{q}_n^2(t)-\frac{\kappa}{2}\sum_{m,n=1}^{\rm N}A_{mn}q_m(t)q_n(t)
\end{equation}
The ion coupling matrix $A_{mn}$ is real, positive and symmetric, which implies it has a number of useful properties.  Its eigenvalues, $\mu_p$, are real and positive; and its eigenvectors, $b^{(p)}_n$, which must be determined numerically except for some simple cases\cite{james_1998}, are orthogonal and complete.  Assuming they are appropriately normalized, the eigenvectors thus have the following properties:
\begin{eqnarray}
    \sum_{m=1}^{\rm N}b^{(p)}_mb^{(q)}_m&=&
    \delta_{pq} \label{eq:bprops1}\\
    \sum_{p=1}^{\rm N}b^{(p)}_mb^{(p)}_n&=&
    \delta_{mn}\label{eq:bprops2}\\
    \sum_{p=1}^{\rm N}\mu_pb^{(p)}_nb^{(p)}_m&=&
    A_{nm}\label{eq:bprops3}
\end{eqnarray}

Using these values, one can define the normal modes of the ion motion,

\begin{equation}\label{eq:normal_modes}
    Q_p(t) = \sum_{m=1}^{\rm N}q_m(t)b^{(p)}_m
\end{equation}

The position of the ions can be written in terms of these modes $q_m(t) = \sum_{p=1}^{\rm N}b^{(p)}_mQ_p(t)$, so the kinetic and potential energy can be written in terms of the normal modes. Therefore, the Lagrangian can be written as,
\begin{equation}\label{eq:LagrangianQ}
    L = \frac{{\rm M}}{2}\sum_{p=1}^{\rm N}\left[\dot{Q}^2_p(t)-\omega_p^2Q_p(t)^2\right].
\end{equation}
Here $\omega_p=\sqrt{\kappa\mu_p/{\rm M}}$ is the frequency of the normal modes. The Euler-Lagrange equation this gives the following simple equation of motion:
\begin{equation}\label{eq:EOM}
   \ddot{Q}_p(t)+\omega_p^2Q_p(t)=0,
\end{equation}
which can be solved straightforwardly to yield
\begin{equation}\label{eq:Qsol}
   Q_p(t)= Q_p(0)\cos(\omega_pt)+\frac{\dot{Q}_p(0)}{\omega_p}\sin(\omega_pt).
\end{equation}

\section{Trapped Ion Collision}
Imagine an ion trap that has been successfully loaded and we have a chain of cold ions in a linear crystal. What would happen if some particle, for example a molecule of the background gas (since no vacuum is ever perfect) collided with one of the ions in the trap?  While the ions can be imaged by scattering of laser light, the particle is invisible: the laser illuminating the ions will be at a non-resonant wavelength.  Can we deduce which ion received the collision?

To answer this question, the motion of the ions needs to be considered. Suppose the collision is an impulse on a single ion, which instantaneously imparts a velocity $v_0$.  The initial position and velocity of the $m$-th ion is given by $q_m(0)=0$, and $\dot{q}_m(0)=\delta_{mn}v_0$, were the $n$-th ion is the one affected by the collision; thus using the definition of the normal modes eq.(\ref{eq:normal_modes}), the initial conditions will be $Q_p(0) = 0$ and $\dot{Q}_p(0)=v_0b^{(p)}_n$.  Using the equations of motion eq.(\ref{eq:Qsol}) and these initial conditions results in the following solution for the normal modes of the ions,  
\begin{equation}\label{eq:mode_solution}
    Q_p(t) = \frac{b^{(p)}_nv_0}{\omega_p}\sin(\omega_pt).
\end{equation}
Thus the position of the $m$-th ion is given by the formula (\ref{eq:mode_solution}).
\begin{equation}
    q_m(t) = \sum_{p=1}\beta^{(p)}_m \sin(\omega_pt)
\end{equation}
where the amplitude associated with each oscillation frequency  is given by
\begin{equation}\label{eq:amplitude}
     \beta^{(p)}_m =v_0 \sqrt{\frac{{\rm M}}{\kappa \mu_p}} b^{(p)}_n b^{(p)}_m.
\end{equation}

Can the value of $n$, which is the number labeling the ion which underwent the collision, be determined from measurements of the spectral components of the motion of one of the ions?  To begin, let us be selective, and chose to measure the motion of the leftmost ion ($m=1$) (although the technique is applicable to position measurements of most of the ions in the chain; note however if ${\rm N}$ is odd, then the central ion  ($m=({\rm N}+1)/2$) should be avoided for this diagnostic, since some modes of this ion will never be excited).

\begin{figure}
    \centering
    \includegraphics[width=0.5\textwidth]{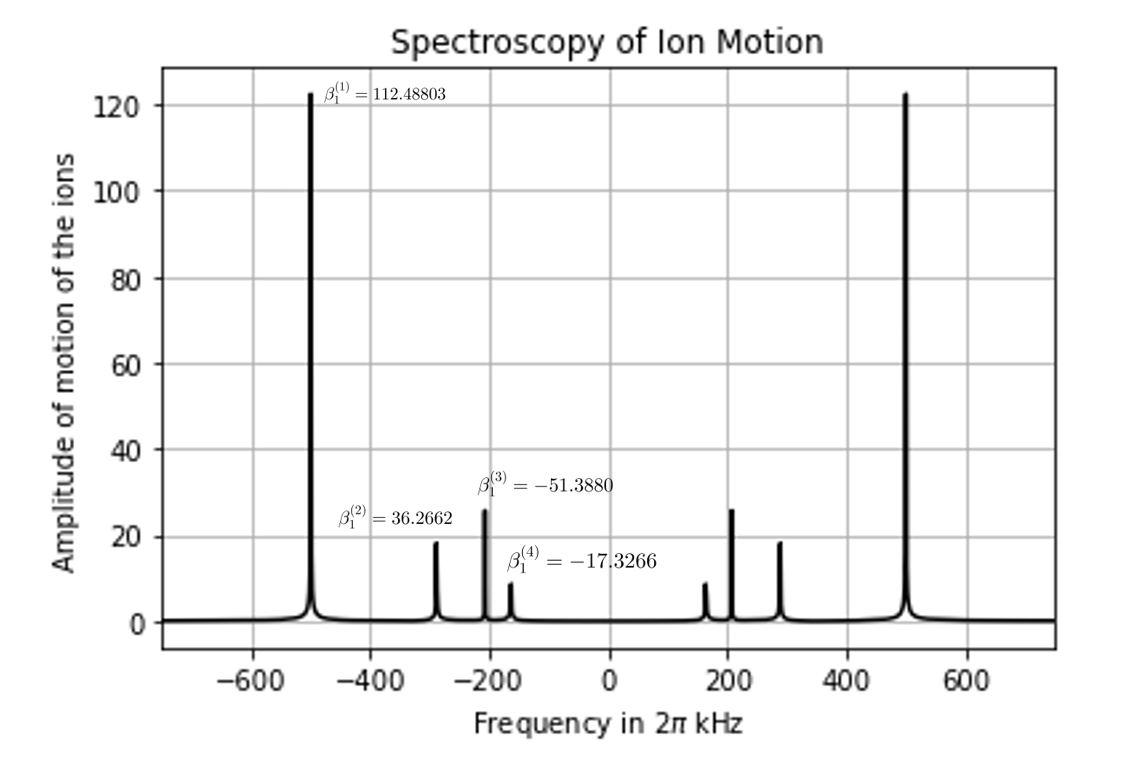}
    \caption{Spectroscopy of the $m=1$ ion showing the amplitude of oscillation ($\beta_m^{(p)}$ of the ions and frequency $\omega_p$ of the oscillations}
    \label{fig:FFT}
\end{figure}

From eq.(\ref{eq:amplitude}), the amplitude for the $m=1$ ion's oscillations in the lowest (Center of Mass) mode, $p=1$, for which $\mu_1=1$, is  
$\beta^{(1)}_1 = v_0 \sqrt{{\rm M}/\kappa}b^{(1)}_n b^{(1)}_1 = v_0/{\rm N} \sqrt{{\rm M}/\kappa }$, where we have used the fact that $b^{(1)}_n=1/\sqrt{{\rm N}}$.  Thus 
\begin{equation}\label{eq:betap1}
     \frac{\beta^{(p)}_1}{\beta^{(1)}_1} = \frac{{\rm N}}{\sqrt{\mu_p}}  b^{(p)}_n b^{(p)}_1.
\end{equation}
Re-arranging, we find
\begin{equation}\label{eq:bpn}
     b^{(p)}_n =\left(\frac{\sqrt{\mu_p}}{{\rm N} b^{(p)}_1}\right) \frac{\beta^{(p)}_1}{\beta^{(1)}_1}.
\end{equation}
All the quantities in the bracket are tabulated (see \cite{james_1998}, Table 2), or can be calculated numerically; the ratio $\beta^{(p)}_1/ \beta^{(1)}_1$ can be obtained from observation, specifically by a spectral analysis of the observed motion of the $m=1$ ion. Figure \ref{fig:FFT} shows a simulation of the Fourier Transform of the motion of the first ion. The amplitudes of each peak are the values for $\beta_1^{(p)}$. Since $\mu_p$, $N$, $b_1^{(p)}$, and $\beta_1^{(1)}$, having the amplitude values allows us to determine the values for $b_n^{(p)}$. We can then use equation \ref{eq:DeltaA} to determine what the value of $n$ is. In the case of Fig.  \ref{fig:FFT}, that results in a value of $n=2$. 

Therefore, if the ratio $\beta_m^{(p)}/\beta_m^{(1)}$ can be determined, the collision site can be determined. 

\section{Unknown Ion in the Chain}
Now let us consider the case of a dark ion in the chain.  Can the ions' dynamics allow us to identity such an impurity?  To address this problem, we must return to the original Lagrangian.  Provided all of the ions, including the impurity, are singly ionized, the potential energy, which depends solely on electric forces, is independent of mass. The kinetic energy is however mass dependent, so the modified Lagrangian is:
\begin{equation}\label{eq:Lagrangianq_mmod}
    L = \frac{1}{2}\sum_{n=1}^{\rm N} {\rm M}_n\dot{q}_n^2(t)-\frac{\kappa}{2}\sum_{m,n=1}^{\rm N}A_{mn}q_m(t)q_n(t)
\end{equation}
where ${\rm M}_n$ is the mass of the $n$-th ion.  Defining a scaled position variable
\begin{equation}
    s_n(t)=\sqrt{\frac{{\rm M}_n}{bar{\rm M}}}q_n(t),
\end{equation}
where $\bar{\rm M}$ is a convenient scale mass, such as the most frequently occurring mass in the chain (i.e. the statistical mode of the mass distribution), the Lagrangian becomes
\begin{equation}
    L = \frac{1}{2}\sum_{n=1}^{\rm N} \bar{\rm M}\dot{s}_n^2(t)-
    \frac{\kappa}{2}\sum_{m,n=1}^{\rm N}
    \frac{\bar{\rm M}}{\sqrt{{\rm M}_n{\rm M}_m}}A_{nm} s_n(t)s_m(t)
\end{equation}
Defining the new ion coupling matrix $A^\prime_{mn}=\bar{\rm M}A_{nm}/\sqrt{{\rm M}_n{\rm M}_m}$, which is real and symmetric, there will be a new orthonormal set of eigenvectors $b_n^{(p)\prime}$ and eigenvalues $\mu^\prime_p$. It follows that the inhomogeneous ion chain will have new normal modes $S_p=\sum_{n=1}^{\rm N} b_n^{(p)\prime} s_n$ with frequencies given by 
$\omega^\prime_p=\sqrt{\kappa\mu^\prime_p/\bar{\rm M}}$. 

Assuming that we can perform a spectroscopic analysis of the motion of the inhomogeneous ion chain, the various mode frequencies can be determined experimentally.  Determining the ion mass distribution in the general case would then be a laborious task of searching through the mode spectra for the possible mass distributions and identifying the one that closest fits the data: there seems no analytic means to perform the inverse problem more directly.  However, if we make the assumption that the variation in mass is always small, a perturbative analysis allows approximate expressions for the unknown mass distribution to be derived from the frequency spectrum.

Let the mass variation be defined by the expression
\begin{equation}
\delta {\rm M}_m={\rm M}_m-\bar{\rm M}.
\end{equation}
We assume $|\delta {\rm M}_m|/\bar{\rm M}\ll 1$, and so the new coupling matrix may be written as $A^\prime_{mn}= A_{mn}+\Delta A_{mn}$ where:
\begin{equation}\label{eq:DeltaA}
\Delta A_{mn}= -\frac{1}{2\bar{\rm M}}\left(\delta {\rm M}_m+\delta {\rm M}_n\right)A_{mn} + O[\delta {\rm M}_m^2].
\end{equation}
Using standard perturbation theory, the eigenvalues of the new
coupling matrix are $\mu^\prime_{p}=\mu_{p}+\Delta\mu_{p}$ where:
\begin{equation}
\Delta\mu_{p} = 
\sum_{m,n=1}^{\rm N} 
b_m^{(p)}\Delta A_{mn} b_n^{(p)} + O[\delta {\rm M}_m^2].
\end{equation}
Substituting from eq.(\ref{eq:DeltaA}), as using the properties of the eigenvectors eqs.(\ref{eq:bprops1})-(\ref{eq:bprops3}), we find
\begin{equation}
\Delta\mu_{p} = -\frac{\mu_p}{\bar{\rm M}}
\sum_{m=1}^{\rm N} 
\left(b_m^{(p)}\right)^2 
\delta {\rm M}_{m} + O[\delta {\rm M}_m^2].
\end{equation}
For the first mode (the ``Center of Mass'' mode, in which all ions oscillate as if firmly clamped together) we have $\mu_1=1$ and $b_m^{(1)}=1/\sqrt{{\rm N}}$; hence
\begin{equation}
\Delta\mu_{1} = -\frac{1}{{\rm N}\bar{\rm M}}
\sum_{m=1}^{\rm N} 
\delta {\rm M}_{m} + O[\delta {\rm M}_m^2].
\end{equation}
%And for the second mode (the ``stretch'' or breathing mode, in which each ion's displacement is proportional to its equilibrium position) we have $\mu_2=3$ and $b_m^{(2)}=x^0_m/\sqrt{\mathcal{N}}$, where $\mathcal{N}=\sum_{m=1}^{\rm N} (x^0_m)^2$; hence
%\begin{equation} \Delta\mu_{2} = -\frac{3 \sum_{m=1}^{\rm N} (x^0_m)^2 \delta {\rm M}_m}{\bar{\rm M} \sum_{m=1}^{\rm N} (x^0_m)^2 }+ O[\delta {\rm M}_m^2].\end{equation}

Since the mode oscillation frequency is $\omega_p^\prime = \sqrt{\kappa \mu_p^\prime /\bar{\rm M}}$, it follows that 
\begin{equation}
\frac{\Delta\omega_p}{\omega_p} 
=\frac{1}{2}\frac{\Delta\mu_p}{\mu_p}.
\end{equation}
Hence, for the ion chain with impurities, the ratio of $p$-th mode frequency to the first mode is
\begin{eqnarray}
\frac{\omega^\prime_p}{\omega^\prime_1} 
&=&\frac{\omega_p+\Delta\omega_p}{\omega_1+\Delta\omega_1}\nonumber\\
&=&\sqrt{\mu_p}
\left(
\frac{
2 \bar{\rm M}-\sum_{m=1}^{\rm N}\delta{\rm M}_m\left(b_m^{(p)}\right)^2
}{
2 \bar{\rm M}-\sum_{m=1}^{\rm N}\delta{\rm M}_m/{\rm N}
}
\right).
\end{eqnarray}
Suppose we have a single impurity ion of unknown mass ${\rm M}_i$ at the $i$-th position in the chain; Thus $\bar{\rm M}={\rm M}$, the mass of the majority of the ions in the chain, and $\delta{\rm M}_m=({\rm M}_i-{\rm M})\delta_{mi}$.  Re-arranging, we obtain the following expression for ${\rm M}_i$:

\begin{equation}
{\rm M}_i = {\rm M}\left[
1+ 2{\rm N}\frac{
\left(\omega^\prime_p/\omega^\prime_1\right)-\sqrt{\mu_p}
}{
\left(\omega^\prime_p/{\rm N}\omega^\prime_1\right)-\sqrt{\mu_p}\left(b_i^{(p)}\right)^2
}.
\right]
\end{equation}

\section{Conclusion}

Despite the many useful applications of trapped ions, they are not perfect. Vacuum's created in lab are never perfect, allowing for different ions to be present in the same space where ion traps are being set up. Collisions against these ions can cause changes in the usually predictable features of the trap\cite{rugango_2015,kimura_2011,fan_2021}. Another issue causing the same effect is when those surrounding ions are trapped instead of the intended ion. These stowaway ions also have effects on the behavior of the trap. To counteract these effect or use them to our advantage the known quantities, such as equations of motion, of the dynamics of the ion need are utilized. 

Using the modes of motion of the ions, the necessary equations to determine the location of a collision and the mass of the stowaway ion. Once a collision site is determined one could remove the ion that was collided against and the trap can resume to its normal motion. On the other hand, determining the mass of the other ion is one step to study reaction rate coefficients of cold chemical reactions, without the requirement of mass spectrometry to determine the masses of the ions in the trap, and for sideband cooling. The latter reducing the effects of trap heating. The equation for the mass of the ion can aid in further understanding of sympathetic cooling and spectroscopy \cite{schmidt_2016, wan_2014}.
%Add citations above.

\section{Acknowledgments}
The authors than Dr. Amar Vutha for useful discussions. Additionally, they would like to thank Clodomiro De Urioste who provided support during all stages of writing and research.

\section*{Disclosure Statement}
The authors report there are no competing interests to declare.

\section*{Funding}
This work was funded by the Natural Sciences and Engineering Council of Canada (RGPIN-2017-06264)

\bibliographystyle{apalike}
\bibliography{references}

\end{document}